\documentclass[twocolumn,preprintnumbers,amsmath,amssymb]{revtex4-1}

\usepackage{graphicx}
\usepackage{dcolumn}
\usepackage{bm}

\usepackage{color}
\definecolor{lightblue}{rgb}{0.2,0.2,0.7}
\definecolor{darkblue}{rgb}{0,0.25,0.5}
\definecolor{redbrown}{rgb}{0.875,0.25,0.125}
\definecolor{darkgreen}{rgb}{0,0.5,0}

\newcommand{\ket}[1]{\ensuremath{\vert #1  \rangle}}

\newcommand{\CSF}{\ensuremath{\text{CSF}}}
\renewcommand{\det}{\ensuremath{\text{det}}}
\newcommand{\exc}{\ensuremath{\text{exc}}}

\begin{document}

\title{Quantum Monte Carlo with Jastrow-Valence-Bond wave functions}

\author{Beno\^it Bra\"ida$^1$}
\email{Author to whom correspondence should be addressed. Electronic mail: benoit.braida@upmc.fr}
\author{Julien Toulouse$^1$}
\author{Michel Caffarel$^2$}
\author{C. J. Umrigar$^3$}
\affiliation{
$^1$Laboratoire de Chimie Th\'eorique, Universit\'e Pierre et Marie Curie and CNRS, Paris, France.\\
$^2$Laboratoire de Chimie et Physique Quantiques, IRSAMC, CNRS and Universit\'e de Toulouse, Toulouse, France.\\
$^3$Laboratory of Atomic and Solid State Physics, Cornell University, Ithaca, New York, USA.
}

\date{\today}
\begin{abstract}
We consider the use in quantum Monte Carlo calculations of two types of valence bond wave functions based on strictly localized active orbitals,
namely valence bond self-consistent-field (VBSCF) and breathing-orbital valence bond (BOVB) wave functions. Complemented by a Jastrow factor,
these Jastrow-Valence-Bond wave functions are tested by computing the equilibrium well depths of the four diatomic molecules C$_2$, N$_2$, O$_2$, and F$_2$
in both variational Monte Carlo (VMC) and diffusion Monte Carlo (DMC). We show that it is possible to design compact wave functions based on chemical grounds
that are capable of describing both static and dynamic electron correlation. These wave functions can be systematically improved by inclusion of
valence bond structures corresponding to additional bonding patterns.

\end{abstract}

\maketitle

\section{Introduction}
\label{sec:intro}

Quantum Monte Carlo (QMC) methods (see e.g. Refs.~\onlinecite{HamLesRey-BOOK-94,NigUmr-BOOK-99,FouMitNeeRaj-RMP-01}) constitute
an alternative to standard quantum chemistry approaches for accurate calculations of the electronic structure of atoms,
molecules and solids. The most commonly-used approach consists of optimizing a flexible trial wave function in a variational
Monte Carlo (VMC) calculation, and then using the resulting wave function in a more accurate fixed-node diffusion Monte Carlo
(DMC) calculation.

For atoms and molecules, the most common form of trial wave function consists of a Jastrow factor, expected to describe a major part of the \textit{dynamic} electron correlation effects, multiplied by a single Slater determinant of orbitals expanded in a localized
one-particle basis. For systems featuring important \textit{static} correlation effects, the single Slater
determinant is usually replaced by a linear combination of several Slater determinants. This has been shown to lead to
systematic improvements of both VMC and DMC total energies, provided that the wave function is properly
reoptimized~\cite{TouUmr-JCP-07,UmrTouFilSorHen-PRL-07,TouUmr-JCP-08}. In particular, full-valence complete active space (FVCAS)
expansions, i.e. including all the Slater determinants that can be generated by distributing all the valence electrons in all
the valence orbitals, were found to yield near chemical accuracy for bonding energies of first-row homonuclear diatomic
molecules at the DMC level~\cite{TouUmr-JCP-08}.

However, FVCAS expansions involve large numbers of Slater determinants that scale exponentially with system size, so this
approach cannot be applied for large systems. Instead, more compact and yet systematic wave functions are desired for QMC
calculations. Naively-truncated expansions of Slater determinants of delocalized molecular orbitals cannot be seen as a
general solution since they often suffer from lack of consistency for calculating energy differences between two systems or two
states. In contrast, pairing-type wave functions should be an adequate, compact and systematic way of describing static
correlation in QMC. Indeed, several forms of pairing wave functions have recently been used in QMC, namely antisymmetrized
geminal power (AGP)~\cite{CasSor-JCP-03,CasAttSor-JCP-04,MarAzaCasSor-JCP-09},
pfaffians~\cite{BajMitDroWag-PRL-06,BajMitWagSch-PRB-08} and generalized valence bond (GVB)~\cite{AndGod-JCP-10}, with promising
results on atoms and small molecules.

Here, we explore the use in QMC of valence bond (VB) wave functions based on nonorthogonal active orbitals that are localized on a single atom (for a review of VB theory, see Refs.~\onlinecite{McW-BOOK-92,Gal-BOOK-02,ShaHib-BOOK-08}).
The merits of valence bond self-consistent field (VBSCF) wave functions~\cite{LenBal-JCP-83},
are tested, along with more flexible breathing-orbital valence bond (BOVB) wave functions~\cite{HibHumByrLen-JCP-94,HibSha-TCA-02}. These approaches can describe static correlation effects with compact wave functions containing only a limited number of VB structures (i.e., linear combinations of Slater determinants made of spin-singlet pairs of orbitals) which are selected on chemical grounds. In comparison to the GVB method, which is usually considered in the perfect-pairing and strong-orthogonality approximations, the VBSCF and BOVB approaches are more flexible and provide a greater chemical interpretability thanks to the use of localized nonorthogonal orbitals. In contrast to traditional analytical calculations, nonorthogonal orbitals do not lead to extra computational cost in QMC.  On the contrary, the use of localized nonorthogonal orbitals could potentially reduce the computational cost as compared with a delocalized molecular-orbital approach in three ways: a) localized orbitals are negligibly small and need not be computed beyond some cutoff distance, b) the number of determinants needed may be reduced by having nonorthogonal orbitals, and, c) in the absence of spatial symmetry, the number of orbital parameters to be optimized is smaller for localized orbitals.

BOVB wave functions improve upon VBSCF wave functions by allowing a different set of orbitals in each VB structure, effectively introducing dynamic
correlation into the Slater determinants, but in a different way than through backflow coordinate
transformations~\cite{LopMaDruTowNee-PRE-06} or orbital-attached multi-Jastrow factors~\cite{BouBraCaf-JCP-10}. In VB practice,
it is known that bonding energies at the BOVB level are significantly improved as compared with the VBSCF level, especially in
cases where dynamic correlation is important. In QMC, this form of wave functions has already successfully been tested on
acetylene dissociation energies~\cite{DomBraLes-JPCA-08}. However, it is not known whether the breathing-orbital relaxation yields significant improvement when the wave function is fully optimized in QMC, because it is conceivable that the Jastrow factor in QMC provides much of the same variational freedom.

The paper is organized as follows. In Section II, we offer a brief review of the VBSCF and BOVB approaches, and describe
our corresponding Jastrow-Valence-Bond wave functions used in QMC. In Section III, we present results for the four diatomic molecules C$_2$, N$_2$, O$_2$, and F$_2$, which span different types of bonding (single, multiple, and three-electron bond) and electron correlation (strong dynamic correlation for O$_2$ and F$_2$, and strong
static correlation for C$_2$ and N$_2$). We discuss the type of electron correlation to include in the wave functions to obtain accurate equilibrium well depths, as well as the importance of orbital optimization in QMC, and the computational cost of our wave functions. Section VI summarizes our conclusions.

\section{Methodology}

\subsection{Valence-Bond wave functions}

The different valence bond methods can be classified into two main families, depending of the degree of localization of the
active orbitals: \textit{semilocalized orbitals} and \textit{strictly localized orbitals}~\cite{HibSha-JComChem-07,
ShaHib-BOOK-08}. In the first family, the optimized active orbitals are expanded on all the basis functions of the entire molecule, and are thus free to fully delocalize, although most of the time they appear to be fairly localized. The GVB method~\cite{BobGod-INC-77,GodDunHunJef-ACR-73} and the spin-coupled
valence bond (SCVB) method~\cite{CooGerRai-CR-91} belong to this family. In the second family, each active orbital is
expanded in basis functions centered on a single atom or the atoms of a fragment of the molecule (alternative definition of {\it strictly} localized orbitals exists, Ref.~\onlinecite{BraTouCafUmr-JJJ-XX-note}). The VBSCF~\cite{LenBal-JCP-83,LenVerPul-MP-91} and
BOVB~\cite{HibHumByrLen-JCP-94,HibSha-TCA-02} methods belong to this family. The semilocalized approaches usually provide more compact wave functions, but the delocalization tails of the orbitals may hinder their chemical readability.
The strictly localized approaches usually involve more VB structures, but permit the distinction between \textit{covalent} and
\textit{ionic} contributions to the bonds, thus providing more insight into the nature of the chemical bonds.
The interpretative power of the strictly localized VB approaches is hence greater, and allows, among other examples, the characterization of charge-shift bonding~\cite{ShaHib-ChemEurJ07,ShaHib-Angew09,ShaHib-NatChem09}, three-electron bonds~\cite{HibHum-JCP-94,BraHib-02,BraHib-04} or hypervalent bonding~\cite{Sini-90,BraHib-04,BraHib-08}. Further, most applications involving VB curve-crossing diagrams~\cite{ShaShu-Angew-99,ShaHib-BOOK-08} are within the strictly localized VB formalism.

Any VBSCF wave function can be written as a linear combination,
\begin{eqnarray}
\ket{\Psi_{\text{VBSCF}}} = \sum_{I} c_I \ket{\Phi_I},
\label{PsiVBSCF}
\end{eqnarray}
where $\ket{\Phi_I}$ are spin-adapted VB \textit{structures}, each one being determined by a choice of an \textit{orbital configuration} (or orbital occupation) and a \textit{spin coupling} of these orbitals. We consider VB structures of the form (disregarding normalization)
\begin{eqnarray}
\ket{\Phi_I} &=& \prod_p^{\text{inactive}} \hat{a}^\dag_{p\uparrow} \hat{a}^\dag_{p\downarrow}
\prod_{(ij)}^{\substack{\text{active}\\\text{pairs}}} \left( \hat{a}^\dag_{i\uparrow} \hat{a}^\dag_{j\downarrow} -
\hat{a}^\dag_{i\downarrow} \hat{a}^\dag_{j\uparrow} \right)
\nonumber\\
&&\times\prod_q^{\substack{\text{active}\\\text{unpaired}}} \hat{a}^\dag_{q\uparrow} \ket{\text{vac}},
\label{PhiI}
\end{eqnarray}
where $\hat{a}^\dag_{p\sigma}$ ($\sigma =\uparrow,\downarrow$) is a spin-orbital creation operator and $\ket{\text{vac}}$ is the
vacuum state of second quantization. The VB structures are thus made of inactive (always closed-shell) orbitals $p$,
spin-singlet pairs of active orbitals $(ij)$, and possibly remaining unpaired spin-up active orbitals $q$. We use inactive
orbitals that are either localized (expanded on the basis functions centered on a single atom), e.g. for core orbitals, or
delocalized (expanded on all the basis functions of all the atoms), e.g. for bonds made of inactive orbitals that do not mix
with the active orbitals. We use active orbitals that are always localized on a single atom, and they are typically identified
with valence atomic (hybrid) orbitals. Note that Eq.~(\ref{PhiI}) encompasses the case of
spin-singlet pairing of an active orbital with itself, i.e. $i=j$ giving simply $\hat{a}^\dag_{i\uparrow}
\hat{a}^\dag_{i\downarrow} - \hat{a}^\dag_{i\downarrow} \hat{a}^\dag_{i\uparrow}=2\hat{a}^\dag_{i\uparrow}
\hat{a}^\dag_{i\downarrow}$. The spin-coupling scheme based on singlet pairing used in Eq.~(\ref{PhiI}) is usually referred to as
the Heitler-London-Slater-Pauling (HLSP) scheme. There exist other spin-coupling schemes, but the HLSP scheme has the advantage
of providing a clear correspondence between each VB structure and a Lewis chemical structure, two singlet-paired active orbitals
representing either a bond or a lone pair. In principle, considering all possible pairings exhausts, for a given orbital
configuration, all the spin eigenstates (or spin couplings) of fixed quantum numbers
$S=N_{\text{unpaired}}/2$ and $M_S=+S$ ($N_{\text{unpaired}}$ is the number of spin-up unpaired electrons). In fact, considering
all possible pairings leads to an overcomplete set of spin couplings, but they can be reduced to a complete basis of
(non-redundant) spin couplings, also called Rumer basis~\cite{ShaHib-BOOK-08,Rumer-32,Pauncz-BOOK-79}.
In practice, for most practical applications, only a small number of chemically relevant VB structures are kept in the calculation.

In practical VBSCF calculations, each VB structure is expanded in $2^{N_{\text{pairs}}}$ Slater determinants, $\ket{\Phi_I}=\sum_{\mu} d_{I,\mu} \ket{D_\mu}$, where
$N_{\text{pairs}}$ is the number of pairs of different active orbitals ($i\not=j$). The coefficients of the determinants $d_{I,\mu}$
for a given VB structure are all equal in absolute value.
The VB structure coefficients and orbital coefficients on the basis are then optimized using a direct generalization of the usual multi-configuration self-consistent-field (MCSCF) algorithm for arbitrary nonorthogonal orbitals~\cite{LenVerPul-MP-91}, or algorithms more specific to VB theory~\cite{WuWuMo-IJQC-98,SongMoWu-JCC-08}. In fact, this VBSCF procedure is general enough to also permit optimization of GVB
and SCVB wave functions if the active orbitals are allowed to delocalize. Note that GVB wave functions, in the most commonly
used perfect-pairing approximation, are made of only one VB structure, while the SCVB wave functions include all possible
spin couplings.

As an illustration, consider the single-bonded Li$_2$ molecule. A typical VBSCF wave function is
\begin{eqnarray}
\ket{\Psi_{\text{VBSCF}}} = c_1 \ket{\Phi_\text{cov}} + c_2 \ket{\Phi_\text{ionA}} + c_3 \ket{\Phi_\text{ionB}},
\label{PsiVBSCFLi2}
\end{eqnarray}
with a covalent VB structure obtained by spin pairing the $2s$ orbital of the first Li (A) with the $2s$ orbital of the second
Li (B)
\begin{eqnarray}
\ket{\Phi_\text{cov}} &=& \hat{a}^\dag_{1s_A\uparrow} \hat{a}^\dag_{1s_A\downarrow} \hat{a}^\dag_{1s_B\uparrow}
\hat{a}^\dag_{1s_B\downarrow}
\nonumber\\
&&\times\left( \hat{a}^\dag_{2s_A\uparrow} \hat{a}^\dag_{2s_B\downarrow} - \hat{a}^\dag_{2s_A\downarrow}
\hat{a}^\dag_{2s_B\uparrow} \right) \ket{\text{vac}},
\label{PhicovLi2}
\end{eqnarray}
and two ionic VB structures obtained by either spin pairing $2s_A$ with itself,
\begin{eqnarray}
\ket{\Phi_\text{ionA}} &=& \hat{a}^\dag_{1s_A\uparrow} \hat{a}^\dag_{1s_A\downarrow} \hat{a}^\dag_{1s_B\uparrow}
\hat{a}^\dag_{1s_B\downarrow} \hat{a}^\dag_{2s_A\uparrow} \hat{a}^\dag_{2s_A\downarrow}  \ket{\text{vac}},
\label{PhiionALi2}
\end{eqnarray}
or, symmetrically, by spin pairing $2s_B$ with itself,
\begin{eqnarray}
\ket{\Phi_\text{ionB}} &=& \hat{a}^\dag_{1s_A\uparrow} \hat{a}^\dag_{1s_A\downarrow} \hat{a}^\dag_{1s_B\uparrow}
\hat{a}^\dag_{1s_B\downarrow} \hat{a}^\dag_{2s_B\uparrow} \hat{a}^\dag_{2s_B\downarrow}  \ket{\text{vac}},
\label{PhiionBLi2}
\end{eqnarray}
the core inactive orbitals $1s_A$ and $1s_B$ remaining always doubly occupied. These three VB structures correspond respectively to three
Lewis chemical structures, Li---Li, $^-$Li\phantom{---}Li$^+$, and $^+$Li\phantom{---}Li$^-$, respectively. Obviously, in this case,
spatial symmetry implies that $c_2=c_3$ and the two ionic VB structures can thus be combined in a single ``symmetry-adapted
structure'', which we refer to as a \textit{configuration state function} (CSF). In this simple case, the VBSCF wave function
is essentially a complete active space (CAS) wave function with the two valence electrons distributed in the
valence bonding $\sigma_g$ and antibonding $\sigma_u$ molecular orbitals, i.e. CAS(2,2), and would be strictly identical to it
in a minimal basis set. In less trivial cases, the selected VB structures correspond only to a subspace of a CAS.
However, if structure selection is correctly made, the VBSCF wave function can lead to observable predictions in close agreement to a CAS wave function.
Indeed, in addition to the most chemically significant structures, the CAS wave function includes a lot of improbable covalent and (multi-)ionic
structures that are of little importance. Thus, a VB wave function can represent an intelligent
way of truncating a CAS expansion based on chemical grounds.

We adopt the usual definitions of electron correlations used in quantum chemistry. Static correlation corresponds to the correlation included in a FVCAS wave function as compared with a Hartree-Fock reference, and dynamic correlation corresponds to the remaining correlation resulting from all the outer-valence excitations in a full configuration interaction expansion. The VBSCF method handles static correlation only. Dynamic correlation can be added in the usual way by performing
configuration interaction on top of VBSCF~\cite{WuSonCaoZhaSha-JPCA-02,SonWuZhaSha-JCC-04}, referred to as VBCI, or
through a perturbative treatment~\cite{SonWuZhaSha-JCC-04,ChenWu-JPCA-09}. The BOVB method is an alternative way of incorporating
some dynamic correlation, keeping the wave function as compact as the VBSCF one. This
method consists in allowing a different set of (inactive and active) orbitals in each VB structure.
The orbitals can thus adapt to the different charge distributions in the VB structures. For example, for the Li$_2$
wave function of Eqs.~(\ref{PsiVBSCFLi2})-(\ref{PhiionBLi2}), in BOVB the $2s$ orbitals can become more diffuse in the
ionic structures than in the covalent structure. This extra degree of freedom is seen as introducing dynamic correlation, since it roughly corresponds to
adding excitations to atomic orbitals of higher principal quantum numbers.
Just as in the VBSCF method, inactive valence orbitals can be chosen to be either
localized or delocalized in the BOVB method. Finally, in the ionic structures,
a doubly-occupied active orbital (the $2s$ orbital for Li$_2$) is sometimes split into two independent orbitals that are
spin-singlet coupled, which introduces additional dynamic correlation.
The most flexible BOVB method is thus obtained by both splitting of active orbitals in the ionic structures
and delocalization of inactive orbitals, which is referred to as split-delocalized-BOVB (SD-BOVB) in the VB literature.

\subsection{Jastrow-Valence-Bond wave functions}

We discuss now the parametrization of the two types of Jastrow-Valence-Bond wave functions, J-VBSCF and J-BOVB, that we use in
QMC. The J-VBSCF wave function is parameterized as (see Ref.~\onlinecite{TouUmr-JCP-08})
\begin{eqnarray}
\ket{\Psi_{\text{J-VBSCF}}} = \hat{J} \, e^{\hat{\kappa}} \sum_{I} c_I \ket{\Phi_I},
\label{PsiJVBSCF}
\end{eqnarray}
where $\hat{J}$ is a Jastrow operator, $e^{\hat{\kappa}}$ is an orbital ``rotation'' operator, and $\ket{\Phi_I}$ are CSFs,
i.e. VB structures or linear combination of a few VB structures to accommodate spatial symmetry. Each CSF is a short
linear combination of products of spin-up and spin-down Slater determinants, composed generally of nonorthogonal orbitals expanded
on Slater basis functions. The orbital operator $\hat{\kappa}$ is $\hat{\kappa} = \sum_{k l} \kappa_{kl} \,  \hat{E}_{kl}$
where $\kappa_{kl}$ are the orbital ``rotation'' parameters and $\hat{E}_{kl}$ is the singlet excitation operator from orbital
$l$ to orbital $k$, $\hat{E}_{kl}=\hat{a}_{k \uparrow}^{\dag} \hat{b}_{l \uparrow} + \hat{a}_{k \downarrow}^\dag \hat{b}_{l
\downarrow}$, written with dual biorthogonal orbital creation and annihilation operators $\hat{a}_{k \sigma}^\dag$ and
$\hat{b}_{l \sigma}$ (see, e.g., Ref.~\onlinecite{HelJorOls-BOOK-02}). The parameters to optimize are the parameters in the
Jastrow factor, the CSF coefficients, the orbital rotation parameters and the exponents of the basis functions.

The J-BOVB wave function is similar except that each CSF is allowed to have a different set of orbitals, which can be written as
\begin{eqnarray}
\ket{\Psi_{\text{J-BOVB}}} = \hat{J} \, \sum_{I} c_I \, e^{\hat{\kappa}_I} \ket{\Phi_I},
\label{PsiJBOVB}
\end{eqnarray}
where $\hat{\kappa}_I$ is the orbital operator for the set of orbitals in the CSF $\ket{\Phi_I}$. In practice,
Eq.~(\ref{PsiJBOVB}) can be recast in the form of Eq.~(\ref{PsiJVBSCF}) by defining the orbital operator $\hat{\kappa}$ to operate on all sets of orbitals:
$\hat{\kappa}=\sum_I \hat{\kappa}_I$.

We use a Jastrow factor consisting of the exponential of the sum of electron-nucleus, electron-electron, and
electron-electron-nucleus terms, written as systematic polynomial and Pad\'e expansions~\cite{Umr-UNP-XX} (see also Refs.
~\onlinecite{FilUmr-JCP-96} and~\onlinecite{GucSanUmrJai-PRB-05}).

\subsection{Computational details}

We start by generating a standard restricted Hartree-Fock (RHF) wave function with the program Gaussian~\cite{Gaussian-PROG-03} and a VBSCF or BOVB wave function with the program XMVB~\cite{SonMoZhaWu-JCC-05}. We use the core-valence triple-zeta quality Slater basis (CVB1) of Ema {\it et al.}~\cite{EmaGarRamLopFerMeiPal-JCC-03}. For the XMVB calculations, each Slater function is
expanded into 15 (core basis functions), 9 ($s$ and $p$ valence basis functions), or 6 ($d$ polarization functions) Gaussian
functions. We use here 5 $d$ spherical functions while 6 $d$ Cartesian functions were used in Ref.~\onlinecite{TouUmr-JCP-08}.
The standard RHF, VBSCF or BOVB wave function is then multiplied by our Jastrow factor, and QMC calculations are performed with
the program CHAMP~\cite{Cha-PROG-XX} using the true Slater basis set rather than its Gaussian expansion.

The Jastrow, CSF, orbital and basis exponent parameters are simultaneously optimized by minimizing the energy plus a small
percentage of the energy variance (1\%) with the linear optimization
method~\cite{TouUmr-JCP-07,UmrTouFilSorHen-PRL-07,TouUmr-JCP-08} in VMC, using an accelerated Metropolis
algorithm~\cite{Umr-PRL-93,Umr-INC-99}. Once the trial wave function has been optimized, we perform a DMC calculation within the
short-time and fixed-node (FN) approximations (see, e.g.,
Refs.~\onlinecite{GriSto-JCP-71,And-JCP-75,And-JCP-76,ReyCepAldLes-JCP-82,MosSchLeeKal-JCP-82}). We use an imaginary time step
of $\tau=0.01$ hartree$^{-1}$ in an efficient DMC algorithm with very small time-step errors~\cite{UmrNigRun-JCP-93}.

For orbital optimization, the orbitals are partitioned into three sets: inactive (doubly occupied in all determinants), active
(occupied in some determinants and unoccupied in others), and virtual (unoccupied in all determinants). All inactive (including 1s core) and active orbitals are optimized. The non-redundant excitations to consider are inactive $\to$ active, inactive $\to$ virtual, active $\to$ virtual and active $\to$ active. If the
action of the excitation $\hat{E}_{kl}$ on the wave function is not zero but the reverse excitation $\hat{E}_{lk}$ is zero, then
the ``orthogonality'' condition $\kappa_{lk} = - \kappa_{kl}$ is imposed. For some active-active excitations, both direct and
reverse excitations ($\hat{E}_{kl}$ and $\hat{E}_{lk}$) may be allowed, and thus it makes sense for localized orbitals to
remove the ``orthogonality'' constraint by treating $\kappa_{kl}$ and $\kappa_{lk}$ as independent parameters. This results in
only a very few (if any at all) additional orbital parameters for the wave functions considered here. We note that, when
considering active-active excitations, redundancies between two orbital wave-function derivatives or between an orbital
wave-function derivative and a CSF frequently occur, and must be detected and eliminated. Localized orbitals do not have the point
group symmetry of the system, so the number of orbital excitations cannot be reduced based on the non mixing of irreducible
representations, as usually done. However, the number of orbital excitations is greatly reduced by forbidding mixing between
orbitals of different localization classes, i.e. expanded on different subsets of basis functions. For BOVB wave
functions, one has a different set of (occupied and unoccupied) orbitals for each CSF~\cite{BraTouCafUmr-JJJ-XX-note2} and only orbitals that belong to the same
set are allowed to mix.

\section{Results}

The Jastrow-Valence-Bond wave functions are tested on the ground states of the molecules C$_2$ ($^1\Sigma_g^+$), N$_2$
($^1\Sigma_g^+$), O$_2$ ($^3\Sigma_g^-$), and F$_2$ ($^1\Sigma_g^+$) at their experimental bond lengths of 1.2425 \AA, 1.0977 \AA,
1.2075 \AA, and 1.4119 \AA, respectively. The core inactive $1s$ orbitals are always taken as localized. For O$_2$ and F$_2$,
we use delocalized inactive $\pi$ orbitals both at the VBSCF and BOVB levels. This reduces the
Pauli repulsion between the otherwise localized lone pairs.
The precise composition of all wave functions is explained below.
In addition to total energies, we present molecular well
depths all computed with respect to the separated atoms calculated at the FVCAS level.
For the N$_2$, O$_2$, and F$_2$ molecules, both VBSCF and BOVB wave functions in the limit of a complete basis of structures (Rumer basis) dissociate into the FVCAS atomic limit, which simply corresponds for these atoms to single-determinant wave functions.
For C$_2$, VBSCF in the limit of a complete basis of structures also dissociates into the FVCAS atomic limit, which corresponds for the carbon atom in its triplet ground state to a CAS(2,2) wave function, since the carbon lone pair can doubly occupy either a $2s$ or a $2p$ orbital.
With this definition, the well depths
actually allow a comparison of the absolute quality of the different molecular wave functions at equilibrium since the atomic reference is the same.

\subsection{Single bonding-pattern wave functions}

\begin{figure}[t]
\includegraphics[scale=0.8,angle=0]{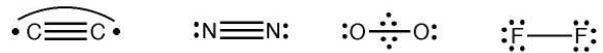}
\caption{Dominant bonding patterns for the C$_2$, N$_2$, O$_2$, and F$_2$ molecules, defining our single bonding-pattern VB wave functions.}
\label{fig:singlespincoupling}
\end{figure}

\begingroup
\squeezetable
\begin{table}[t]
\caption{Relative contributions of the different types of electron correlation to the equilibrium well depths for C$_2$, N$_2$, O$_2$, and F$_2$. Starting from the RHF well depth, the ``left-right'' static correlation is defined as the contribution recovered by a single-bonding pattern VBSCF calculation, the other parts of static correlation as the additional contributions recovered by a FVCAS MCSCF calculation, and the dynamic correlation as the remaining contribution necessary to reach the exact well depth.
}
\begin{tabular}[b]{l c c c}
\hline\hline
                              & \multicolumn{2}{c}{Static correlation}  & \multicolumn{1}{c}{Dynamic correlation}    \\
                                \cline{2-3}
                              & \multicolumn{1}{c}{left-right} & \multicolumn{1}{c}{other} & \\
\hline
C$_2$                         & 77\%  & 18\%  & 5\% \\
N$_2$                         & 70\%  & 14\%  & 16\% \\
O$_2$                         & 60\%  & 8\%   & 32\% \\
F$_2$                         & 60\%  & 8\%   & 32\% \\
\hline\hline
\end{tabular}
\label{tab:correlation}
\end{table}
\endgroup

\begingroup
\squeezetable
\begin{table*}[t]
\caption{Total energies (in hartree) and well depths (in kcal/mol) for the ground states of O$_2$ and F$_2$ at their experimental bond lengths calculated in RHF, VBSCF and BOVB, and in VMC and DMC using corresponding Jastrow-Valence-Bond wave functions with all the parameters (Jastrow, CSF coefficients, orbitals and basis exponents) fully optimized in VMC. For comparison, DMC results for FVCAS wave functions from a previous study are also shown.}
\begin{tabular}[b]{l l l l l}
\hline\hline
                              & \multicolumn{2}{c}{O$_2$}                & \multicolumn{2}{c}{F$_2$} \\
                              & \multicolumn{1}{c}{Total energies}  & \multicolumn{1}{c}{Well depths} & \multicolumn{1}{c}{Total energies}  & \multicolumn{1}{c}{Well depths} \\
\hline
RHF                           &   -149.6567        &     22.4            &  -198.7608           &   -36.2         \\
VBSCF                         &   -149.7512        &     81.7            &  -198.8327           &     8.9         \\
BOVB$^c$                      &   -149.8042$^d$    &     114.9$^d$       &  -198.8627           &    27.8         \\
BOVB-split$^e$                &                    &                     &  -198.8689           &    31.7         \\
\\
VMC J-RHF                     &   -150.2287(5)     &     99.0(4)         &  -199.4204(5)        &    12.4(4)      \\
VMC J-VBSCF                   &   -150.2414(5)     &    107.0(4)         &  -199.4400(5)        &    24.7(4)      \\
VMC J-BOVB$^c$                &   -150.2496(5)     &    112.1(4)         &  -199.4456(5)        &    28.3(4)      \\
VMC J-BOVB-split$^e$          &                    &                     &  -199.4494(5)        &    30.6(4)     \\
\\
DMC J-RHF                     &   -150.2861(2)     &    114.7(2)         &  -199.4833(2)        &    29.8(2)      \\
DMC J-VBSCF                   &   -150.2889(2)     &    116.5(2)         &  -199.4900(2)        &    34.0(2)      \\
DMC J-BOVB$^c$                &   -150.2922(2)     &    118.5(2)         &  -199.4922(2)        &    35.4(2)      \\
DMC J-BOVB-split$^e$          &                    &                     &  -199.4962(2)         &    37.9(2)     \\
\\
DMC J-FVCAS$^a$               &   -150.2944(1)     &    119.6(1)         &  -199.4970(1)        &    37.9(1)      \\
Estimated exact$^b$           &   -150.3274        &   120.9($<$1)         &  -199.5304           &    39.0(1)      \\
\hline\hline
\end{tabular}\\
\begin{flushleft}
\footnotesize
$^a$ Ref.~\onlinecite{TouUmr-JCP-08}.\\
$^b$ Ref.~\onlinecite{BytRue-JCP-05}.\\
$^c$ BOVB with delocalized $\pi$ orbitals, which is referred to as $\pi$-D-BOVB in the VB literature.\\
$^d$ Starting from an optimized BOVB wave function with all localized orbitals, delocalized $\pi$ orbitals are allowed and optimized but keeping all the other orbitals frozen. Optimization of all the orbitals with delocalized $\pi$ orbitals was not stable. This problem was not encountered in the VMC optimization.\\
$^e$ ``BOVB-split'' (usually referred to as SD-BOVB in the VB literature): in each of the ionic structures, the doubly occupied active orbital is split into two singlet-coupled singly occupied orbitals.
\end{flushleft}
\label{tab:o2_f2}
\end{table*}
\endgroup

We first consider wave functions describing what we call a single \textit{bonding pattern}. A bonding pattern is a compact Lewis-style picture of the molecular electronic structure, but which is made of several (covalent and ionic) VB structures. The dominant bonding patterns of C$_2$, N$_2$, O$_2$, and F$_2$ are depicted in Fig.~\ref{fig:singlespincoupling}. The conventions of the drawings are the following. Each single line connecting two atoms represents a $\sigma$ or $\pi$ two-electron bond. In our wave functions, each two-electron bond is generally composed of one covalent and two ionic components. Hence, F$_2$ (single bond) is described by a wave function containing 3 VB structures, whereas N$_2$ (triple bond) leads to a wave function with $3^3=27$ VB structures. Each pair of dots on a single atom represents a lone pair, which is described by two opposite-spin electrons occupying the same orbital.

The single dots connected by a curve which appears for C$_2$ represent a (formally covalent) singlet coupling between two
electrons occupying two different orbitals, one located on each carbon in that case. The two corresponding coupled orbitals are
$sp$ hybrids pointing in opposite direction, so that they overlap and thus interact very weakly. Hence, this corresponds to a
singlet biradical state, the coupling between the two single electrons being only formal to ensure the proper spin symmetry.
Very recent VB calculations by Su \textit{et al.}~\cite{WuShaHib-11} have shown this biradical triple-bond situation to be the dominant bonding pattern for C$_2$ at equilibrium distance. In order to keep the number of VB structures low, only the single-ionic structures, as well as the double-ionic structures with opposite charges on each carbon, are included in our wave function. This leads to a wave function with 21 VB structures.

The O$_2$ molecule is considered in its spin-triplet (paramagnetic) ground state. The picture in Fig.~\ref{fig:singlespincoupling}  corresponds to the
textbook molecular-orbital diagram for this molecule, displaying a $\sigma$ two-electron bond (single line connecting the two oxygens), and two
$\pi$ three-electron half-bonds, each one represented by a three-dot symbol. In our wave functions, the $\sigma$ bond is
composed of one covalent and two ionic components, and each three-electron bond is composed of the two resonating structures
typical of this kind of bond~\cite{HibHum-JCP-94,BraHib-02}. This leads to a wave function with 12 VB structures.

In each wave function, structures equivalent by symmetry are further grouped together into CSFs. For each dimer, the VBSCF and BOVB wave functions have exactly the same number of CSFs, the only difference being that different sets of orbitals are used for the different CSFs of the BOVB wave functions.

A rough analysis of the relative contributions of the different types of electron correlation to the equilibrium well depths for the four molecules is given in Table~\ref{tab:correlation}. This analysis leads us to consider separately the case of the O$_2$ and F$_2$ molecules which have strong dynamic correlation, and the case of the C$_2$ and N$_2$ molecules which have strong static correlation.

\subsubsection{Systems with strong dynamic correlation: O$_2$ and F$_2$}

The total energies and well depths of O$_2$ and F$_2$ computed with standard VB methods and QMC methods are displayed in Table~\ref{tab:o2_f2}.
We start by analyzing the standard VB results. The difference between the RHF and VBSCF energies can be taken as the definition of the ``left-right'' part of static correlation.
In VB theory, the left-right correlation is viewed as the energy gained by restoration of the
correct balance between covalent and ionic contributions to the bond, starting from the incorrect 50\% covalent - 50\% ionic
RHF description. The contributions of left-right correlation to the well depths appear to be very large here, 59.3 and 45.1 kcal/mol for O$_2$ and F$_2$,
respectively. A significant part of dynamic correlation is retrieved by going from VBSCF to BOVB,
improving well depths by 33.2 and 18.9 kcal/mol for O$_2$ and F$_2$, respectively. Another 3.9 kcal/mol improvement is obtained for F$_2$ by splitting
the doubly-occupied active orbital in the ionic structures, which brings radial correlation to the active electron pair.
The agreement between the BOVB and exact well depths is quite reasonable considering the compactness of these wave functions: only 12 VB structures (combined into 5 CSFs) for O$_2$ and 3 VB structures (combined into 2 CSFs) for F$_2$.

We now discuss the VMC results. The Jastrow factor keeps electrons apart and describes part of dynamic and static correlation. It is a particularly efficient way of including dynamic correlation, that otherwise requires in configuration interaction calculations a very large number of highly excited determinants. It also includes some static left-right correlation since it greatly reduces the probability of ionic configurations~\cite{TouUmr-JCP-08}.
As expected, a sharp decrease of the total energies is observed in VMC, indicating that a large amount of the \textit{total} correlation energy is recovered by the Jastrow factor. However, for well depths, and most chemical properties, we are instead interested in a balanced description of the \textit{differential} correlation energy in the molecule and the separated atoms. The well depths improve considerably going from RHF to J-RHF, by 76.6 and 48.6 kcal/mol for O$_2$ and F$_2$, respectively, confirming that the Jastrow factor recovers as well a significant part of left-right static correlation. Consequently, going from J-RHF to J-VBSCF wave functions improves the VMC well depths by only 8.0 and 11.7 kcal/mol for O$_2$ and F$_2$, respectively, much less that is gained by going from standard RHF to standard VBSCF. The J-VBSCF VMC well depths remain in relatively poor agreement with the exact values, underestimated by 13.9 and 14.3 kcal/mol for O$_2$ and F$_2$, respectively. It is expected that these errors are mainly due to some missing dynamic correlation, rather than static correlation corresponding to some missing VB structures. This point will be checked in Section III.B.

Additional correlation can be introduced in VMC by replacing the VBSCF determinant expansion by the more flexible BOVB form.
By using a different set of orbitals in each VB structure, the BOVB wave function allows the orbitals to be more diffuse
on the negatively charged atom, and more compact on the positively charged atom of the ionic structures, compared to the neutral covalent structure.
This ``breathing-orbital effect'' can be seen as a manifestation of dynamic correlation.
It is clear that such breathing-orbital effects are partly redundant with the correlations introduced by the Jastrow factor. In practice, it is observed that with the limited form of the Jastrow factor employed, the J-BOVB wave functions improve a bit upon J-VBSCF wave functions: the VMC well depths are bettered by 5.1 and 2.5 kcal/mol
for O$_2$ and F$_2$, respectively.
As expected, these improvements are much smaller than those obtained when going from standard VBSCF to BOVB,
showing that a large part of dynamic correlation is already described by the Jastrow factor.
The agreement between the VMC J-BOVB well depths and the exact values is still not good, with an underestimation of 8.8 kcal/mol for O$_2$, and 10.7 kcal/mol (8.4 kcal/mol with BOVB-split) for F$_2$, respectively.
Even more disappointing, the VMC J-BOVB well depth is actually \textit{worse} than the standard BOVB one for O$_2$, and the same is true for BOVB-split for F$_2$. This indicates that our Jastrow factor describes correlation effects in the atom more accurately than in the (less symmetric) diatomic molecule.
By incorporating more correlation in the separated atoms than in the molecule, the Jastrow factor thus makes the BOVB well depths less accurate.
In fact, it is significant that for F$_2$ splitting the ionic electron pairs improves the VMC well depth by 2.3 kcal/mol,
indicating that this radial correlation effect in the molecule was not well described by the Jastrow factor. The more flexible multi-Jastrow approach of Ref.~\onlinecite{BouBraCaf-JCP-10} could solve this problem by better handling the delicate balance between atomic and molecular correlation effects~\cite{BraTouCafUmr-JJJ-XX-note3}.

Let us now consider the fixed-node DMC results. The total energies and the well depths reflect the quality of the nodes of the different wave functions.
Both total energies and well depths are much improved compared to the VMC results. J-RHF wave functions give DMC well depths of moderate accuracy,
underestimated by 6.2 and 9.2 kcal/mol for O$_2$ and F$_2$, respectively. J-VBSCF wave functions improves DMC well depths but by rather limited amounts, 1.8 and 4.2 kcal/mol for O$_2$ and F$_2$, respectively, showing that most of left-right correlation was already recovered with the J-RHF nodes. Using J-BOVB wave functions allow further gains in accuracy, leading to results close to the ones obtained with much less compact J-FVCAS wave functions.
The best DMC J-BOVB calculations underestimate the well depths by only 2.4 and 1.1 kcal/mol for O$_2$ and F$_2$, respectively.

\subsubsection{Systems with strong static correlation: C$_2$ and N$_2$}

\begingroup
\squeezetable
\begin{table*}[t]
\caption{Total energies (in hartree) and well depths (in kcal/mol) for the ground states of C$_2$ and N$_2$ at their experimental bond lengths calculated in RHF and VBSCF, and in VMC and DMC using corresponding Jastrow-Valence-Bond wave functions with all the parameters (Jastrow, CSF coefficients, orbitals and basis exponents) fully optimized in VMC. For comparison, DMC results for FVCAS wave functions from a previous study are also shown.}
\begin{tabular}[b]{l l l l l}
\hline\hline
                              & \multicolumn{2}{c}{C$_2$}               & \multicolumn{2}{c}{N$_2$} \\
                              & \multicolumn{1}{c}{Total energies}  & \multicolumn{1}{c}{Well depths} & \multicolumn{1}{c}{Total
energies}  & \multicolumn{1}{c}{Well depths} \\
\hline
RHF                           &  -75.4014          &    -8.8             &  -108.9856              &   115.8        \\
VBSCF                         &  -75.5936          &   111.8             &  -109.1112              &   194.6        \\
\\
VMC J-RHF                     &  -75.8132(5)       &   101.9(4)          &  -109.4516(5)           &   204.2(4)     \\
VMC J-VBSCF                   &  -75.8635(5)       &   133.4(4)          &  -109.4723(5)           &   217.2(4)     \\
\\
DMC J-RHF                     &  -75.8667(2)       &   121.8(2)          &  -109.5039(2)           &   221.2(2)      \\
DMC J-VBSCF                   &  -75.8954(2)       &   139.9(2)          &  -109.5119(2)           &   226.3(2)      \\
\\
DMC J-FVCAS$^a$               &  -75.9106(1)       &   149.5(1)          &  -109.5206(1)           &   231.5(1)      \\
Estimated exact$^b$           &  -75.9265          &   148.5(5)          &  -109.5427              &   228.5($<$1)  \\
\hline\hline
\end{tabular}
\begin{flushleft}
\footnotesize
$^a$ Ref.~\onlinecite{TouUmr-JCP-08}.\\
$^b$ Ref.~\onlinecite{BytRue-JCP-05}.
\end{flushleft}
\label{tab:c2_n2}
\end{table*}
\endgroup

Let us now consider C$_2$ and N$_2$, two molecules having strong static correlation even at equilibrium distance. RHF is
expected to be particularly poor for these molecules, because of the triple bond nature of these systems.
In a RHF wave function, each two-electron bond is constrained to be 50\%
covalent - 50\% ionic, and in addition all (multi-)ionic forms are constrained to have the same weights, which is unphysical.
In particular, bi-ionic structures with two positive charges on one atom and two
negative charges on the other will have the same weight as neutral bi-ionic structures (one positive and one negative charge on each atom).
Similarly, tri-ionic structures with three (positive or negative) charges of the same sign on each atom will be forced to have the same
weight as the mono-charged-atom tri-ionic structures.
Note that GVB wave functions also have these constraints on the weights of multi-ionic structures, this form of VB wave function being only able to
take into account intrapair left-right correlation (covalent vs. ionic balance within only one two-electron bond) but not interpair correlation.
On the other hand, a VBSCF wave function can include both intrapair and interpair left-right correlation.
More importantly, for C$_2$, the singlet biradical character is expected to generate additional static correlation beyond left-right correlation, which will require extra bonding patterns.

The total energies and well depths of these molecules are displayed in Table~\ref{tab:c2_n2}.
As expected, RHF strongly underestimates the well depths (note that the negative well depth obtained for C$_2$ is in fact due to the choice of the
FVCAS reference for the separate atoms). VBSCF appears to retrieve a large part of the missing correlation, as it improves
the well depths by 120.6 and 78.8 kcal/mol for C$_2$ and N$_2$, respectively, as compared with RHF values. The VBSCF well depths are however still
underestimated by more than 30 kcal/mol for both molecules. It was impossible to carry out standard BOVB calculations to assess dynamic correlation effects at the VB level, as the number of VB structures (21 for C$_2$ and 27 for N$_2$) is too large to allow such calculations.
Last, we checked that GVB (not shown in Table~\ref{tab:c2_n2}) greatly underestimates the well depths as expected, giving 84.0 and 164.9 kcal/mol
for C$_2$ and N$_2$, respectively, values which are about 30 kcal/mol worse than those from VBSCF.

As for O$_2$ and F$_2$, addition of a Jastrow factor in VMC considerably improves the results.
This improvement is dramatic for single-determinant wave functions, the J-RHF well depths being bettered by
110.7 and 88.4 kcal/mol for C$_2$ and N$_2$, respectively, as compared with the RHF ones.
The improvement is much less important but still significant for the VBSCF wave functions, with J-VBSCF well depths
improved by 21.6 and 22.6 kcal/mol for C$_2$ and N$_2$, respectively, compared to the VBSCF ones.
As for O$_2$ and F$_2$, these findings show that the Jastrow factor can retrieve a significant amount of left-right correlation.
The accuracy of the VMC J-VBSCF calculations is similar to what has been found for O$_2$ and F$_2$, with well depths underestimated by 15.1 and 11.3 kcal/mol
for C$_2$ and N$_2$, respectively.

The DMC results follow the trends found in VMC, but the differences between C$_2$ and N$_2$ show up
more clearly. The N$_2$ well depth is underestimated by 7.3 kcal/mol with a J-RHF wave function, which is actually in the range of what
was observed for O$_2$ and F$_2$. A J-VBSCF wave function leads to a DMC well depth close to chemical accuracy,
indicating that the quality of the J-VBSCF nodes are good enough  get good energy differences.
The situation is quite different for C$_2$.
The J-RHF DMC well depth is underestimated by 26.7 kcal/mol, far from the accuracy reached for the other molecules.
Using a J-VBSCF wave function greatly improves the DMC well depth, but with an error of 8.6 kcal/mol, it is still far from chemical accuracy.
For O$_2$ and F$_2$, a further improvement is possible by using J-BOVB wave functions, but for C$_2$ even if it were possible
to use a J-BOVB wave function a significant improvement is not expected. Because of the multiradical character of C$_2$, we
expect the missing correlation to still be of static nature, rather than dynamic.
The route for improvement is to go beyond a single bonding-pattern description.

\subsection{Multiple bonding-pattern wave functions}

\begin{figure}[t]
\includegraphics[scale=0.8,angle=0]{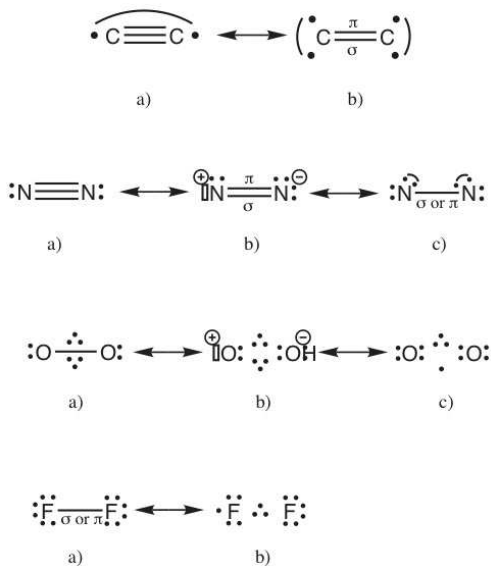}
\caption{Most important resonant bonding patterns for the C$_2$, N$_2$, O$_2$, and F$_2$ molecules, defining our multiple bonding-pattern VB wave functions.}
\label{fig:multiplebonding}
\end{figure}

A straightforward way of improving our J-VBSCF wave functions, particularly for the C$_2$ molecule, is to increase the number of VB structures
included in the wave function. Adding VB structures should be done in a systematic but selective way, as the total
number of non-redundant VB structures (for all orbital configurations and spin couplings) can be very large, e.g. up to 1764 structures for C$_2$ (for the formula giving the number of VB structures in a complete (Rumer) basis, see Ref.~\onlinecite{Weyl-56}).
A route to extend the VB description beyond a single bonding pattern is to systematically derive possible alternative bonding patterns, and
classify them in terms of stability following Lewis-type rules. In particular, the most bonded patterns will be preferred over the
less bonded ones, a $\sigma$ bond will be preferred over a $\pi$ one, and a three-electron (or one-electron) bond will be counted as
half a bond. Following these rules of thumb, a description extended to multiple bonding patterns is proposed in Fig.~\ref{fig:multiplebonding}
for the C$_2$, N$_2$, O$_2$, and F$_2$ molecules.

The main bonding pattern for C$_2$ is of acetylenic type, displaying one $\sigma$ and two $\pi$ bonds, as well as two
singlet-coupled (biradical) electrons. It is natural to postulate for this molecule a secondary bonding pattern of
ethylenic type, with one $\sigma$ and one $\pi$ bond, thus leaving one pair of singlet-coupled electrons on each atom. Another
doubly bonded pattern could be postulated for C$_2$, with two $\pi$ bonds. However, by applying our simple rules, it
is expected to be less stable than the ethylenic pattern, as $\sigma$ bonds are usually more stable than $\pi$ ones. In a recent
VB study, this $\pi$-doubly bonded pattern has indeed been found to be of very little importance for the ground state at
equilibrium distance~\cite{WuShaHib-11}, so it will not be considered here. Considering a C$_2$ description with two bonding
patterns (acetylenic and ethylenic), and including the two covalent and all mono-ionic and neutral bi-ionic structures,
leads to a set of 35 VB structures, which can be combined into only 12 CSFs when all the structures that must have equal weights by symmetry are
combined. All calculated well depths are gathered in Table~\ref{tab:energies}. The first three lines just repeat, for ease of comparison, the well depths obtained using
single bonding-pattern (J-)VBSCF wave functions, previously shown in Tables~\ref{tab:o2_f2} and~\ref{tab:c2_n2}. The
next three lines display the well depths obtained using multiple bonding-patterns (J-)VBSCF wave functions, and the last three lines
some reference data. At the standard VBSCF level already, the well depth of C$_2$ is significantly improved, by 15.4 kcal/mol, when going from
the single bonding-pattern to the multiple bonding-pattern wave function.
Interestingly, a large part of this gain is retained at the VMC J-VBSCF level, with a 9 kcal/mol improvement.
This suggests that the Jastrow function is not able to significantly recover the part of static correlation that is
included when going from single bonding pattern to multiple bonding patterns.
This observation partly transfers to the DMC results as well, with a significant 4.9 kcal/mol improvement of the well depth.
In conclusion, the multiple bonding-pattern J-VBSCF wave function for C$_2$ gives both VMC and DMC well depths in reasonable agreement with estimated exact values
(errors of 6.1 and 3.7 kcal/mol, respectively), while still being a very compact wave function (only 12 CSFs compared to 165 CSFs for the FVCAS wave function).
The remaining error could be progressively eliminated by adding VB structures corresponding to additional bonding patterns, starting with the pattern with two $\pi$ bonds and no $\sigma$ bond, and then including patterns with only one ($\sigma$ or $\pi$) bond. It is expected that a J-VBSCF wave function including all the main VB structures belonging to these bonding patterns would give almost exact dissociation energies, as has recently been shown to be the case at the VBCI level~\cite{WuShaHib-11}, while still having a significantly more compact wave function than the FVCAS one.

\begingroup
\squeezetable
\begin{table}[t]
\caption{
Well depths (in kcal/mol) for the ground states of C$_2$, N$_2$, O$_2$ and F$_2$ at their experimental bond lengths calculated
in VBSCF, VMC J-VBSCF and DMC J-VBSCF using single and multiple bonding-pattern wave functions. All the parameters of the J-VBSCF wave functions (Jastrow, CSF coefficients, orbitals and basis exponents) have been fully optimized in VMC. For comparison, results for FVCAS wave functions from a previous study are also shown.}
\begin{tabular}[b]{l l l l l}
\hline\hline
                              & \multicolumn{1}{c}{C$_2$} & \multicolumn{1}{c}{N$_2$} & \multicolumn{1}{c}{O$_2$} &
\multicolumn{1}{c}{F$_2$}\\
\hline
                              & \multicolumn{4}{c}{\textit{Single bonding pattern}}\\
VBSCF                         &   111.8             &   194.6             &     81.7             &     8.9   \\
VMC J-VBSCF                   &   133.4(4)              &   217.2(4)              &    107.0(4)              &    24.7(4)    \\
DMC J-VBSCF                   &   139.9(2)              &   226.3(2)              &    116.5(2)              &    34.0(2)    \\
\\
                              &  \multicolumn{4}{c}{\textit{Multiple bonding patterns}}\\
VBSCF                         &   127.2             &   203.4             &     84.1             &     9.6   \\
VMC J-VBSCF                   &   142.4(4)              &   220.2(4)              &    108.2(4)              &    26.9(4)   \\
DMC J-VBSCF                   &   144.8(2)              &   228.1(2)              &    117.5(2)              &    35.7(2)   \\
\\
                               & \multicolumn{4}{c}{\textit{References}}\\
MCSCF FVCAS$^a$                &  140.8                  &   210.0                  &    89.9                 &    15.3       \\
VMC J-FVCAS$^a$                &  146.9(1)               &   225.5(2)               &   108.7(2)              &    27.4(2)    \\
DMC J-FVCAS$^a$                &  149.5(1)               &   231.5(1)              &   119.6(1)              &    37.9(1)
\\
Estimated exact$^b$            &  148.5(5)               &   228.5($<$1)           &   120.9($<$1)             &    39.0(1)
\\
\hline\hline
\end{tabular}
\begin{flushleft}
\footnotesize
$^a$ Ref.~\onlinecite{TouUmr-JCP-08}.\\
$^b$ Ref.~\onlinecite{BytRue-JCP-05}.
\end{flushleft}
\label{tab:energies}
\end{table}
\endgroup

\begingroup
\squeezetable
\begin{table*}[t]
\caption{
Well depths (in kcal/mol) for the ground states of C$_2$, N$_2$, O$_2$ and F$_2$ at their experimental bond lengths calculated
in VMC and DMC using single bonding-pattern J-VBSCF wave functions with partial or complete VMC optimization of the parameters: Jastrow (J),
CSF coefficients (c), orbital coefficients (o) and basis exponents (e).
}
\begin{tabular}[b]{l l l l l l}
\hline\hline
               & optimized param.              & \multicolumn{1}{c}{C$_2$} & \multicolumn{1}{c}{N$_2$} & \multicolumn{1}{c}{O$_2$} &
\multicolumn{1}{c}{F$_2$}\\
\hline
VMC J-RHF   & J               &    86.4(4)              &   191.0(4)              &    84.5(4)              &    -0.2(4)    \\
VMC J-RHF   & J+o             &   100.9(4)              &   202.7(4)              &    91.0(4)              &    11.7(4)    \\
VMC J-RHF   & J+o+e           &   101.9(4)              &   204.2(4)              &    99.0(4)              &    12.4(4)    \\
VMC J-VBSCF & J+c             &   125.6(4)              &   207.6(4)              &    95.8(4)              &    13.4(4)    \\
VMC J-VBSCF & J+c+o           &   129.5(4)              &   213.8(4)              &    100.0(4)             &    24.0(4)    \\
VMC J-VBSCF & J+c+o+e         &   133.4(4)              &   217.2(4)              &    107.0(4)             &    24.7(4)     \\
\\
DMC J-RHF   & J                &   114.5(2)              &   218.8(2)              &    110.9(2)             &     25.0(2)   \\
DMC J-RHF   & J+o              &   122.1(2)              &   219.7(2)              &    113.0(2)             &     30.8(2)   \\
DMC J-RHF   & J+o+e            &   121.8(2)              &   221.2(2)              &    114.7(2)             &     29.8(2)   \\
DMC J-VBSCF & J+c              &   135.5(2)              &   222.7(2)              &    113.8(2)             &     30.1(2)    \\
DMC J-VBSCF & J+c+o            &   138.5(2)              &   224.4(2)              &    113.6(2)             &     33.4(2)    \\
DMC J-VBSCF & J+c+o+e          &   139.9(2)              &   226.3(2)              &    116.5(2)             &    34.0(2)     \\
\\
Estimated exact$^a$ &          &  148.5(5)               &   228.5($<$1)           &   120.9($<$1)             &    39.0(1)
\\
\hline\hline
\end{tabular}
\begin{flushleft}
\footnotesize
$^a$ Ref.~\onlinecite{BytRue-JCP-05}.
\end{flushleft}
\label{tab:orbopt}
\end{table*}
\endgroup

Following the same reasoning as for C$_2$, a secondary bonding pattern displaying one $\sigma$ and one $\pi$ bond can be
postulated for the N$_2$ molecule, corresponding to picture b in Fig.~\ref{fig:multiplebonding}. This bonding pattern could also be obtained by
starting from the common Lewis structure of spin-singlet O$_2$, and doubly ionizing one of the two lone pairs.
For N$_2$, a $\pi$-doubly bonded pattern is likely to be of even less importance
than for C$_2$, as this would result in an N atom with a completely filled $\sigma$ space, i.e. two (highly repulsive) $\sigma$ lone
pairs facing each other, so we rule it out. Last, singly bonded patterns with either a $\sigma$ or a $\pi$ bond, corresponding to picture c, are also included,
by analogy with the secondary bonding pattern of C$_2$.
For all these extra patterns, the covalent and all the neutral and mono-charged-atom ionic VB structures are included.
This gives 27 extra VB structures, and so a total of 54 structures (with the main bonding pattern) which can be combined into 24 CSFs.
This multiple bonding-pattern wave function improves the VBSCF well depth of N$_2$ by 15.4 kcal/mol in comparison to the single bonding-pattern calculation,
while still being a substantially more compact wave function than the FVCAS one (24 vs. 107 CSFs).
The improvements in VMC and DMC are less pronounced, though not negligible, with a 2-3 kcal/mol gain.
The DMC J-VBSCF well depth is very close to the estimated exact value, indicating the good quality of the nodes of the multiple bonding-pattern wave function.

For O$_2$, the main bonding pattern corresponds to a double-bond situation, with a $\sigma$ bond and two $\pi$ half bonds.
There are not many possibilities for deriving alternative patterns that are compatible with the spatial and spin-triplet symmetry of the ground
state, and in particular it is impossible to preserve the $\sigma$ bond.
Pattern b keeps the two three-electron bonds, while pattern c has one three-electron and
one one-electron $\pi$ half-bonds and a completely filled $\sigma$ space on each atom. There are a total number of 12 VB structures emerging from
patterns b and c, and when added to the previous set we arrive at a total number of 24 structures (10 CSFs). Obviously,
even if these extra patterns can be associated with a formal bond order of one, they are not expected to be of much importance
in the ground state. This expectation is already confirmed at the standard VBSCF level, as the well depth is improved by only
2.4 kcal/mol, much less than what was observed for C$_2$ and N$_2$. Quite logically, the gains are
even smaller in VMC and DMC, with about 1 kcal/mol improvement only.
It thus seems that for O$_2$ the single bonding-pattern wave function already contains most of the static correlation, and that further gain
in accuracy requires instead improvement of the description of dynamic correlation, as previously
found when going from J-VBSCF to J-BOVB wave functions.

For F$_2$, alternatives to the usual $\sigma$ bond situation are limited too. A $\pi$ bond situation can be
postulated, however it comes with a (highly repulsive) completely filled $\sigma$ space on each atom. A situation with a $\sigma$
three-electron bond can be envisaged (picture b) as it corresponds to the (quite stable) ground state of the difluorine
anion, but it comes here with a singly-ionized $\sigma$ lone pair. Considering all these bonding patterns generates a set
of 13 VB structures (5 CSFs). As expected, the multiple bonding-pattern wave function gives very small improvements.
As for O$_2$, a much more significant improvement was previously obtained for F$_2$ by using a BOVB wave function.

\begingroup
\squeezetable
\begin{table*}[t]
\caption{Numbers of CSFs ($N_{\CSF}$), numbers of unique spin-up + spin-down determinants ($N_\det$) in ground-state wave
functions and numbers of unique spin-up + spin-down singly excited determinants ($N^\exc_\det$) for orbital optimization for the
single bonding-pattern (SBP) and multiple bonding-pattern (MBP) VBSCF and BOVB wave functions employed in this work. Values for FVCAS
wave functions are also shown for comparison.}
\begin{tabular}[b]{l  r r r  r r r  r r r  r r r}
\hline\hline
           & \multicolumn{3}{c}{C$_2$}            & \multicolumn{3}{c}{N$_2$} & \multicolumn{3}{c}{O$_2$} & \multicolumn{3}{c}{F$_2$}\\
\hline
          & $N_{\CSF}$ & $N_\det$& $N^\exc_\det$ & $N_{\CSF}$ & $N_\det$ & $N^\exc_\det$ & $N_{\CSF}$&$N_\det$ &$N^\exc_\det$ &
$N_{\CSF}$& $N_\det$ &$N^\exc_\det$\\
SBP VBSCF &   8 &  32 & 1488 &    10 & 16  & 1280 &    5 & 10 &  848 &    2 & 4  &  232 \\
SBP BOVB  &     &     &      &       &     &      &    5 & 32 & 2144 &    2 & 20 & 1428  \\
MBP VBSCF &  12 &  64 & 2992 &    24 & 72  & 5152 &   10 & 28 & 2282 &    5 & 16 & 1572  \\
FVCAS     & 165 & 140 & 6120 &   107 & 112 & 5864 &   30 & 52 & 2548 &    8 & 16 & 984   \\

\hline\hline
\end{tabular}
\label{tab:cost}
\end{table*}
\endgroup

\subsection{Effect of orbital optimization}

Although for some systems reoptimizing the orbitals in QMC in the presence of the Jastrow factor leads to only small improvements, it is known that it can be very important for other systems, e.g. for calculating certain excitation energies~\cite{SchFil-JCP-04,SceFil-PRB-06,ZimTouZhaMusUmr-JCP-09}. To examine the effect of orbital optimization, we show in Table~\ref{tab:orbopt} the VMC and DMC well depths of the four molecules for J-RHF and single bonding-pattern J-VBSCF wave functions with or without VMC optimization of the orbital coefficients and the basis exponents. In each case, of course, the same level of optimization has been used for the atoms. At the VMC level, optimization of the orbital coefficients and the basis exponents improves the well depths by the order of 10 kcal/mol for J-RHF and J-VBSCF wave functions. In DMC, the gain from optimization of the orbital coefficients and the basis exponents is smaller but still significant, ranging from about 2 to 7 kcal/mol. This shows that orbital optimization in QMC has a significant impact of the nodes of the wave functions. These improvements in VMC and DMC are primarily due to the optimization of the orbital coefficients, the gains from reoptimization of the basis exponents being generally smaller, except in the case of O$_2$.

\subsection{Computational cost}

Finally, we discuss the computational cost of our Jastrow-Valence-Bond wave functions. In QMC calculations, using multideterminant wave functions, the major bottleneck is the evaluation of Slater determinants along with their first- and second-order derivatives with respect to electron coordinates, so that the computational cost is usually directly determined by the number of these determinants. In addition, orbital optimization in VMC using a super-configuration-interaction-type algorithm~\cite{TouUmr-JCP-07,TouUmr-JCP-08} requires evaluation of numerous singly excited determinants, which is often the bottleneck of the optimization calculations. It is thus desirable to have compact wave functions with small numbers of determinants. In practice, each determinant is actually broken into the product of a spin-up and a spin-down determinant, and only the list of unique spin-up and spin-down determinants is computed. In Table~\ref{tab:cost}, we have reported the numbers of CSFs, numbers of unique spin-up + spin-down determinants, and the numbers of unique spin-up + spin-down singly excited determinants for the VBSCF and BOVB wave functions employed in this work. It is seen that there are about 4-7 times less determinants to evaluate for the single bonding-pattern VBSCF wave functions, in comparison to the FVCAS wave functions. For orbital optimization, these single bonding-pattern VBSCF wave functions also generate about 3-5 times less singly excited determinants. On the other hand, single bonding-pattern BOVB wave functions do not appear as computationally advantageous in comparison to FVCAS wave functions. For O$_2$, the BOVB wave function has only a few less determinants, and for F$_2$ the BOVB wave function in fact has more determinants than the FVCAS wave function. Multiple bonding-pattern VBSCF wave functions give less determinants than FVCAS wave functions for C$_2$, N$_2$, and O$_2$, but the reduction is rather modest with at most about a factor of $2$ less determinants for C$_2$. For F$_2$, the multiple bonding-pattern VBSCF wave function actually generates more singly excited determinants than the FVCAS wave function. This happens because for these small F$_2$ wave functions, it is more advantageous to restrict the number of single excitations based on the spatial symmetry of the molecular orbitals, as done in the FVCAS calculations, rather than based on a localization criterion of the VBSCF orbitals.

\section{Conclusions}

We have performed VMC and DMC calculations using compact Jastrow-Valence-Bond wave functions based on strictly localized active orbitals.
The J-VBSCF wave functions are based on a chemically intuitive description of bonding, including either a single bonding pattern (Fig.~\ref{fig:singlespincoupling}) or multiple bonding patterns (Fig.~\ref{fig:multiplebonding}). The J-BOVB wave functions introduce additional correlation through orbital relaxation.

For N$_2$, O$_2$, and F$_2$, the single bonding-pattern J-VBSCF wave functions yield DMC equilibrium well depths with errors ranging from about 2 to 5 kcal/mol. This is quite a reasonable accuracy considering the compactness of the wave functions. For C$_2$, a system with very strong static correlation, a two-bonding-pattern J-VBSCF wave function is necessary to reach a similar accuracy. Optimization of all the wave-function parameters in VMC, and in particular the orbitals, is important to reach this accuracy. As regards computational cost, these VBSCF wave functions contain significantly less Slater determinants than FVCAS expansions, and are therefore more efficient to evaluate. The J-BOVB wave functions are much more costly to evaluate than the J-VBSCF wave functions, and give only moderate improvement of well depths.

The VB {\it ansatz} provides a general way to design compact wave functions that capture only the essential physics of the chemical bond, by including only the most chemically relevant structures. These wave functions could be further improved, by including more and more structures belonging to secondary bonding patterns, depending on the accuracy needed.

For first-row diatomic molecules this approach is of limited interest since it is easily possible to use FVCAS expansions. However, for molecules requiring
 larger active spaces, very compact valence bond wave functions can still be designed, and this becomes a great advantage over molecular orbital based multi-determinant approaches.
Also, for large systems with regions of interest localized on a few atoms, which is usually the case in chemical reactions, strictly localized orbitals enable both a reduction of the active space and a reduction of the orbital optimization cost, as compared to delocalized molecular orbital approaches.

\begin{acknowledgments}
BB thanks Philippe C. Hiberty for helpful discussions and Arne L\"uchow for comments which helped improve the manuscript. CJU acknowledges support from NSF grant number CHE-1004603.
\end{acknowledgments}

\newpage

\bibliographystyle{apsrev4-1}

\begin{thebibliography}{67}
\bibitem[{Bra({\natexlab{a}})}]{BraTouCafUmr-JJJ-XX-note}
\bibinfo{note}{In the condensed-matter community, ``strictly localized
  orbitals'' often refers to orbitals that vanish exactly at some finite
  distance. In the present work, ``strictly localized orbitals'' is employed in
  the sense usually used in the quantum chemistry community, i.e. orbitals
  expanded on Gaussian or Slater basis functions centered on a single atom.
  Thus, these orbitals vanish exactly only at infinite distance.}
\bibitem[{Bra({\natexlab{b}})}]{BraTouCafUmr-JJJ-XX-note2}
\bibinfo{note}{In our BOVB wave functions, we use different sets of orbital
  coefficients, but a single set of basis exponents shared by all orbital
  sets.}
\end{thebibliography}

\end{document}